\documentclass{appolb}
\usepackage{graphicx}
\usepackage{amsmath}
\usepackage{amssymb}
\usepackage{cancel}
\usepackage{hyperref}
\usepackage{cite}
\usepackage{subfig}
\usepackage{xspace}

\newcommand{\ord}{{\ensuremath{\cal O}}}

\newcommand{\ppb}{pP\lowercase{b}\xspace}

\begin{document}
\title{LHC lead data and nuclear PDFs%
\thanks{Presented by A.~Kusina at the Cracow Epiphany Conference on Particle Theory Meets the First Data from LHC Run2,
9-12 January 2017, Krak\'ow, Poland.}%
}
\author{A.~Kusina$^{1,2}$, F.~Lyonnet$^{3}$, D.~B.~Clark$^{3}$, E.~Godat$^{3}$,
        T.~Je\v{z}o$^{4}$, K.~Kova\v{r}\'{\i}k$^{5}$, F.~I.~Olness$^{3}$, I.~Schienbein$^{1}$, J.~Y.~Yu$^{3}$
\address{
$^1$ Laboratoire de Physique Subatomique et de Cosmologie\\ 
     53 Rue des Martyrs Grenoble, France
\\$^2$ Institute of Nuclear Physics, Polish Academy of Sciences,\\
       ul.\ Radzikowskiego 152, 31-342 Cracow, Poland
\\$^3$ Southern Methodist University, Dallas, TX 75275, USA
\\$^{4}$ Physik-Institut, Universit{\"a}t Z{\"u}rich, Winterthurerstrasse 190, CH-8057 Z{\"u}rich, Switzerland
\\$^{5}$ Institut f{\"u}r Theoretische Physik, Westf{\"a}lische Wilhelms-Universit{\"a}t M{\"u}nster,\\
         Wilhelm-Klemm-Stra{\ss}e 9, D-48149 M{\"u}nster, Germany}
}
\maketitle
\begin{abstract}
%
We compare predictions of nCTEQ15 nuclear parton distribution functions
with proton-lead vector boson production data from the LHC. We select
data sets that are most sensitive to nuclear PDFs and have potential to
constrain them. We identify the kinematic regions and flavours where
these data can bring new information and will have largest impact on the
nuclear PDFs. Finally, we estimate the effect of including these data
in a global analysis using a reweighting method.
%
\end{abstract}
%

\section{Introduction}
\label{sec:intro}
\begin{figure}
\centering{}
\includegraphics[width=0.45\textwidth]{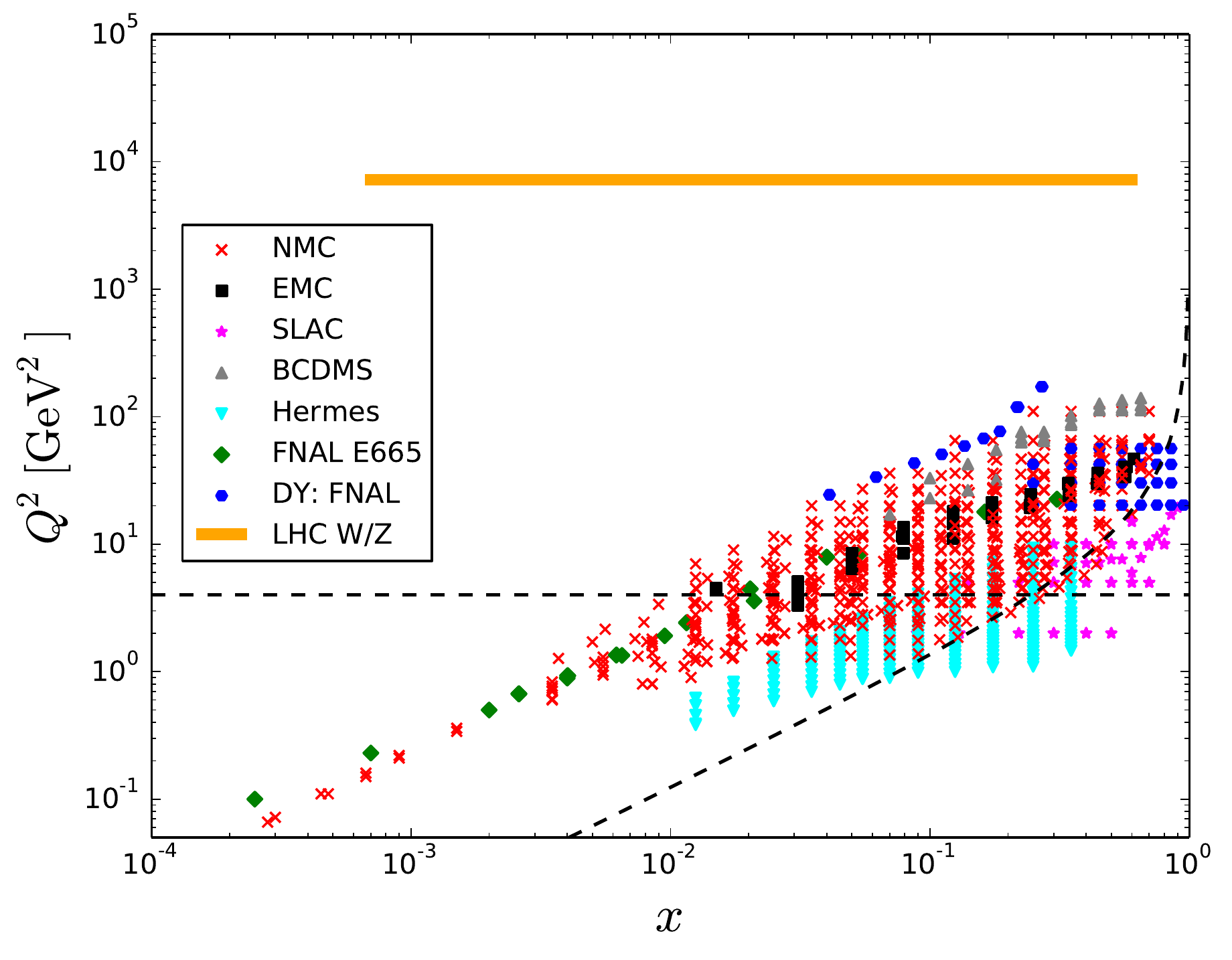}
\hfil
\includegraphics[width=0.45\textwidth]{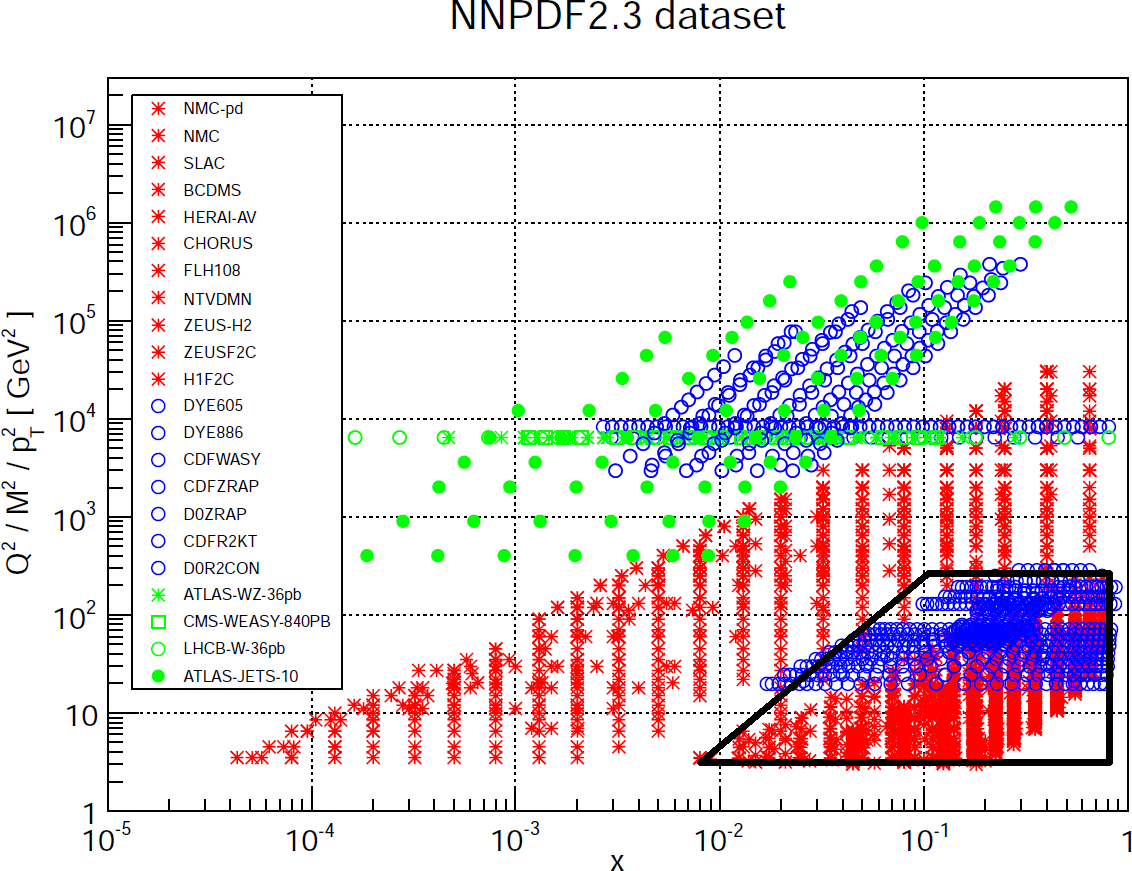}
\caption{(left) Typical kinematical range of data used in nPDF
global analysis with addition of $W/Z$ pPb LHC data.
(right) Example kinematical range of data used in free-proton PDF
global analysis~\cite{Ball:2012cx}.}
\label{fig:kienamtics}
\end{figure}

\begin{figure}
\centering{}
\hspace{-0.75cm}
\includegraphics[width=1.0\textwidth]{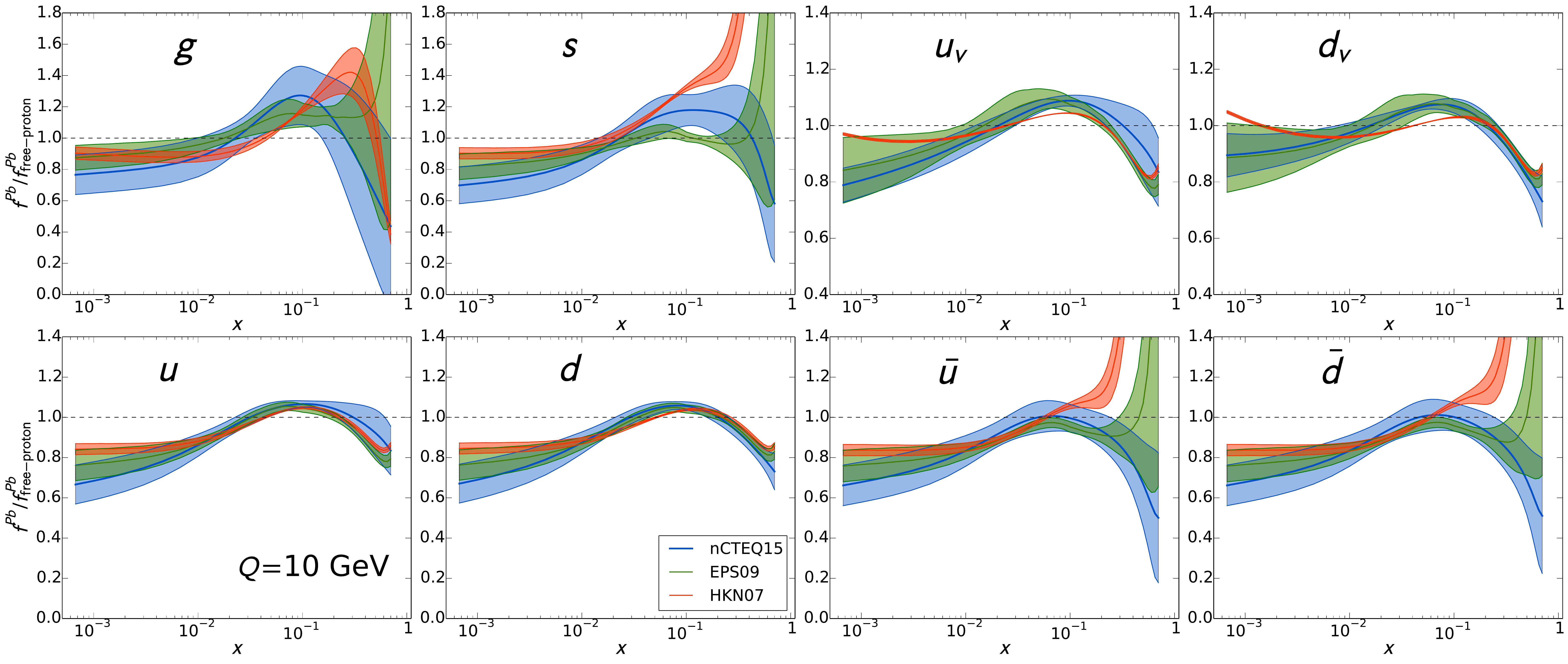}
\caption{Comparison of lead nuclear modifications, $R^{Pb}_i=\frac{f^{Pb}}{f^{Pb}_{\text{free-proton}}}$,
obtained by nCTEQ15~\cite{Kovarik:2015cma}, EPS09~\cite{Eskola:2009uj} and HKN07~\cite{Hirai:2007sx}
nPDF global analyses. Figure from~\cite{Kovarik:2015cma}.}
\label{fig:PDFs-compare}
\end{figure}

Nuclear parton distribution functions (nPDFs) are important quantities
necessary to describe high energy collisions including heavy ions,
as well as giving insight into the structure of nuclei.
nPDFs are non-perturbative objects that can not be calculated by the known
methods. %
Instead, similarly to what is done in the case of proton PDFs, nPDFs are extracted
from experimental data in the process of global analysis. However, in the nPDF case
not only the $x$-dependence is modeled, but also the $A$-dependence (where $A$ is the nucleus mass).
This is partly by design to have a general parametrization of different nuclei,
and partly by necessity as there is typically not a sufficient amount of experimental
data to constrain distributions for individual nuclei separately.
The lack of kinematical data is actually one of the main practical differences between
the proton and nuclear PDF fits. 
This is illustrated in Fig.~\ref{fig:kienamtics} where we compare the kinematical range
of data in both cases.
The lack of data is also the reason why in many cases additional assumptions need to
be introduced in the nPDF analyses in order to obtain stable fits. This necessity,
however, can lead to sizable differences (much bigger than for proton PDFs) between
different nPDFs, see Fig.~\ref{fig:PDFs-compare}.

Vector boson production in hadron collisions is a very well understood
process and serves as one of the  ``standard candle'' measurements at
the LHC. $W^{\pm}$ and $Z$ bosons are numerously produced in heavy ion
proton-lead (pPb) and lead-lead (PbPb) collisions at the LHC and can be
used to gain insight into the structure of nPDFs.
As the $W$ and $Z$ bosons couple weakly, their interaction with the nuclear
medium is negligible which makes these processes one of the cleanest probes
of the nuclear structure available at the LHC. The possibility of using vector
boson production data to constrain nPDFs was considered before the LHC data
were available~\cite{Paukkunen:2010qg}, and this demonstrated a strong
potential for the proton-lead data to constrain nPDFs. 
The current LHC measurements for  $W^{\pm}$ and $Z$ production include 
mostly rapidity distributions for both pPb and PbPb
collisions~\cite{Aad:2015gta,Khachatryan:2015pzs,Aaij:2014pvu,Khachatryan:2015hha,AtlasWpPb,Senosi:2015omk,Aad:2012ew,Chatrchyan:2014csa,Aad:2014bha,Chatrchyan:2012nt}.
Some of these data were already used in a reweighting analyses~\cite{Armesto:2015lrg}
and more recently~\cite{Kusina:2016fxy} to estimate the impact of these data on
EPS09~\cite{Eskola:2009uj}, DSSZ~\cite{deFlorian:2011fp} and nCTEQ15~\cite{Kovarik:2015cma} nPDFs.
Analysis of these data was also performed within the framework of KP model~\cite{Ru:2016wfx}.
Lately, a first global analysis of nPDFs with LHC data, EPPS16~\cite{Eskola:2016oht},
have been published.%
    \footnote{Interesting scaling properties of the LHC $W$ production data from
    pp, pPb and PbPb collisions have been observed in~\cite{Arleo:2015dba}.}

In this work we present predictions for vector boson production in
pPb and PbPb collisions at the LHC obtained using nCTEQ15 nuclear
parton distributions, and perform a comprehensive comparison
to the available LHC data. We also identify the measurements which
have the biggest potential to constrain the nPDFs with special attention
to the strange distribution. Finally, we perform a reweighting
study which gives indications on the effects of including these data in
the nCTEQ global fit. 
This proceedings is based on the recent study~\cite{Kusina:2016fxy} with
additional material on the nuclear strange distribution.

\section{Comparison to the LHC vector boson data}
\label{sec:compar}
We start by comparing predictions for vector boson production at the LHC
calculated using nCTEQ15 nPDFs~\cite{Kovarik:2015cma} to the available
experimental data (for the proton beam we use CT10 proton PDFs~\cite{Lai:2010vv}).
In this note we concentrate only on the most relevant
data sets, namely $W^{\pm}$ production in pPb collisions ($\sqrt{s}=5.02$ TeV)
from CMS~\cite{Khachatryan:2015hha} and ATLAS~\cite{AtlasWpPb}.
A more comprehensive comparison with all the currently available vector
boson data can be found in ref.~\cite{Kusina:2016fxy}.
Our calculations are done using next-to-leading order (NLO) with help of
FEWZ~\cite{Gavin:2012sy} program.

\begin{figure}
\centering{}
\subfloat[CMS $W^{+}$\label{fig:cms_wpm_pPb_comp_wp}]{
\includegraphics[width=0.48\textwidth]{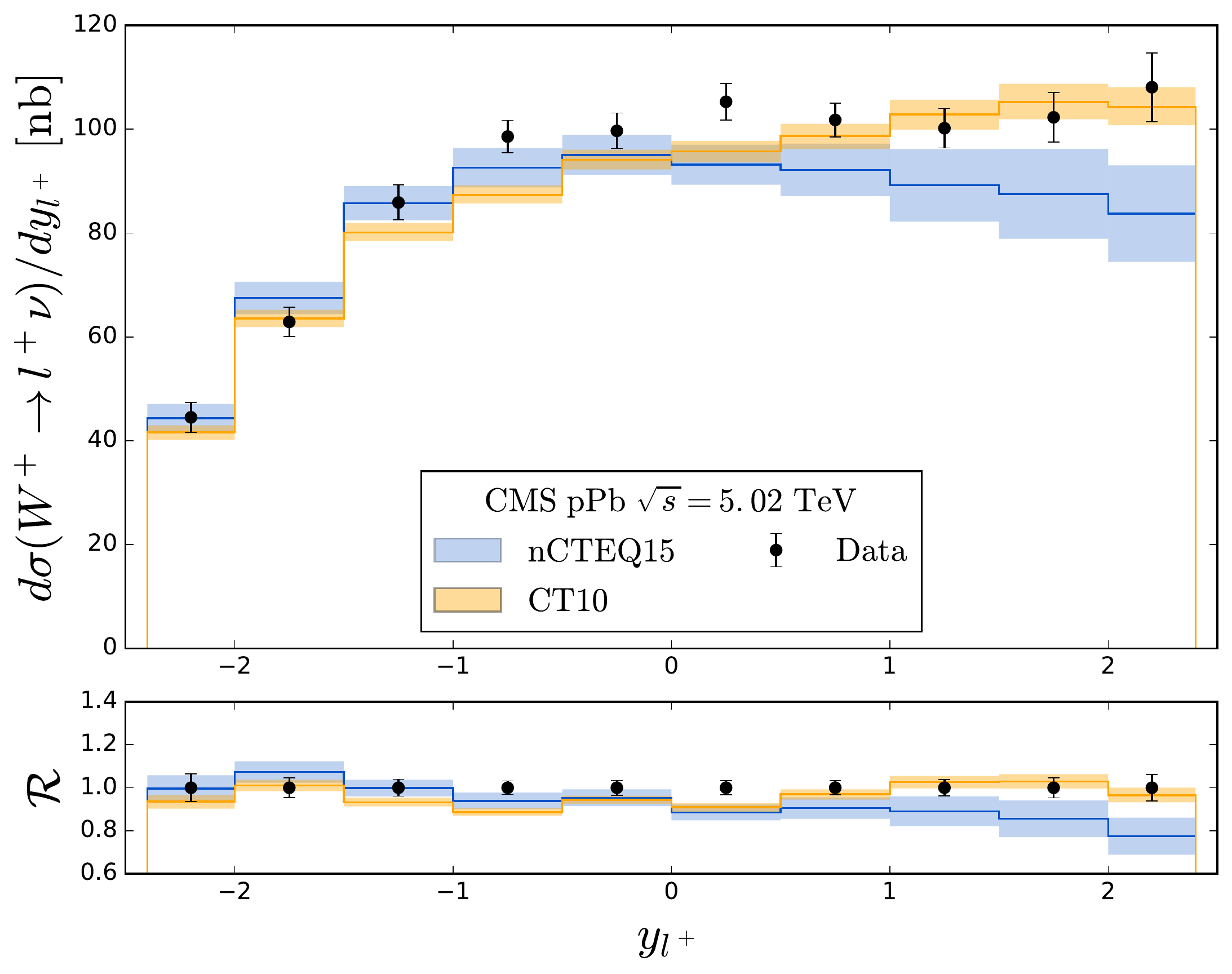}}
\hfil
\subfloat[CMS $W^{-}$\label{fig:cms_wpm_pPb_comp_wm}]{
\includegraphics[width=0.48\textwidth]{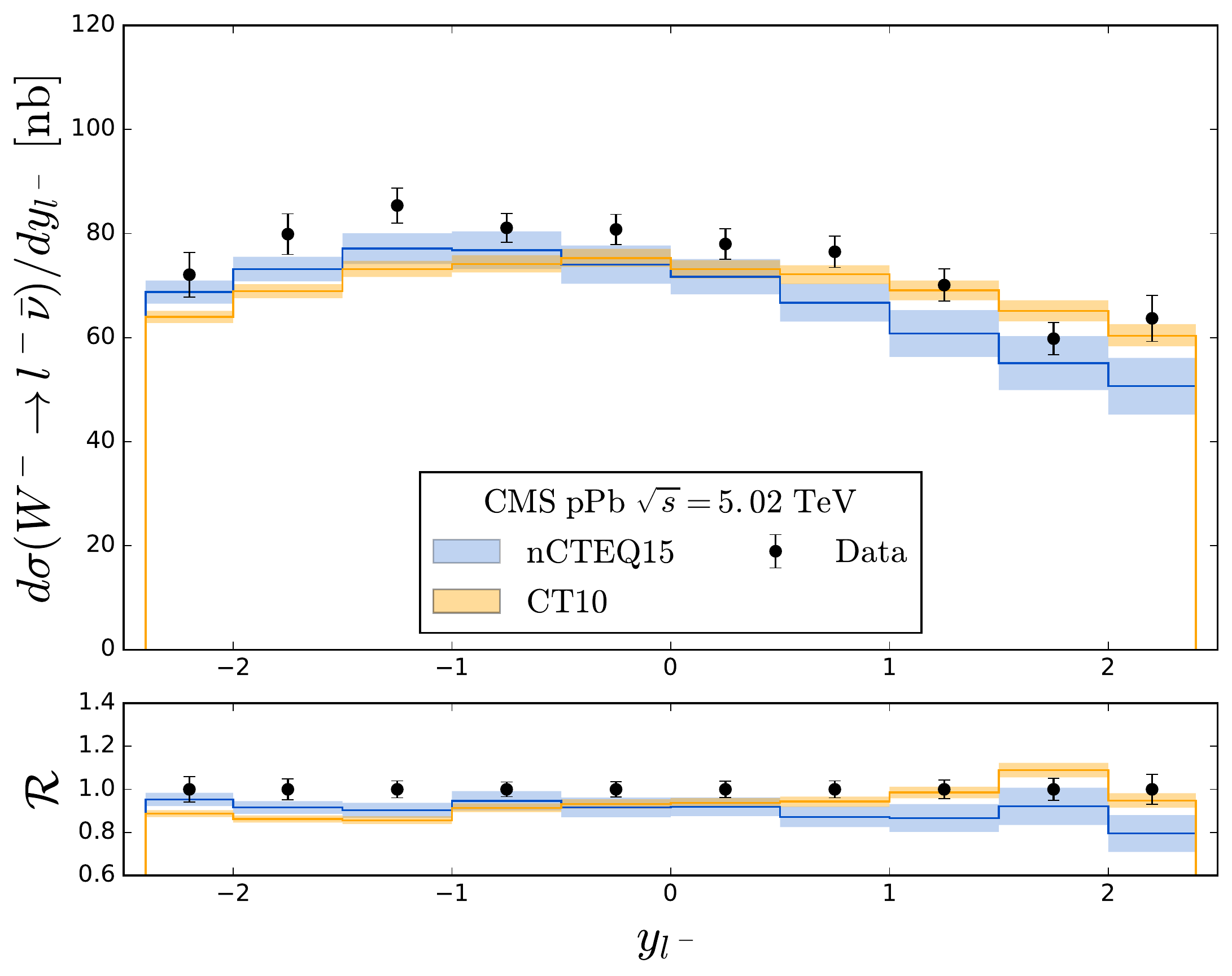}}
\\
\subfloat[ATLAS $W^{+}$\label{fig:atlas_wpm_pPb_comp_wp}]{
\includegraphics[width=0.48\textwidth]{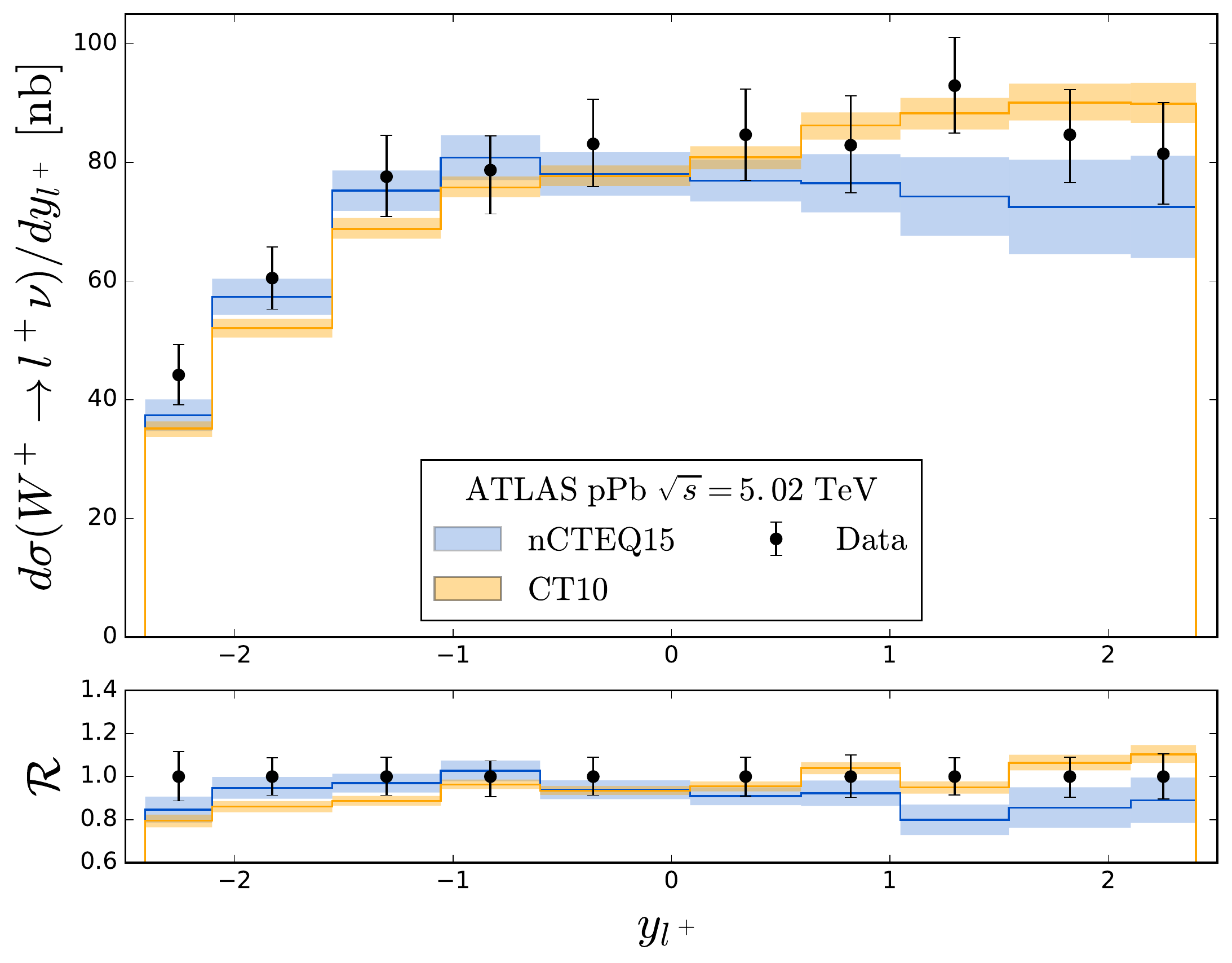}}
\hfil
\subfloat[ATLAS $W^{-}$\label{fig:atlas_wpm_pPb_comp_wm}]{
\includegraphics[width=0.48\textwidth]{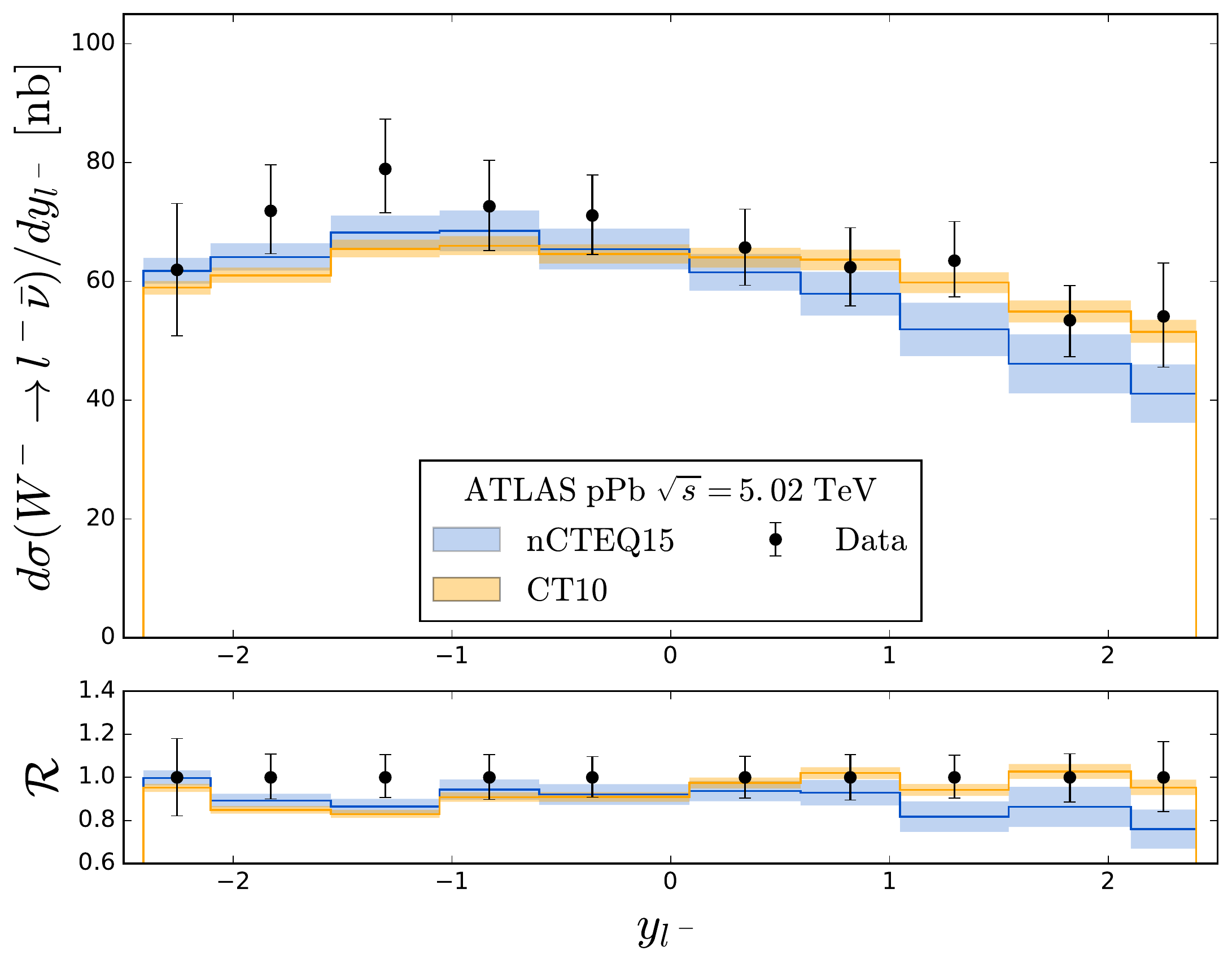}}
\caption{$W^{\pm}$ production in pPb collisions at the LHC from
CMS (upper plots) and ATLAS (lower plots) compared with predictions from
nCTEQ15 nPDFs and CT10 proton PDFs.}
\label{fig:CMS-ATLAS_wpm}
\end{figure}

In Fig.~\ref{fig:CMS-ATLAS_wpm} we present results for
CMS~\cite{Khachatryan:2015hha} and ATLAS~\cite{AtlasWpPb} data.
In the plots we show data overlaid with predictions using nCTEQ15 nPDFs
(blue band) and additionally we show results obtained with free proton PDFs
(yellow band) for which we choose CT10 distributions~\cite{Lai:2010vv}.%
    \footnote{We include here the isospin effects.}

\begin{figure}
\centering{}
\includegraphics[width=0.7\textwidth]{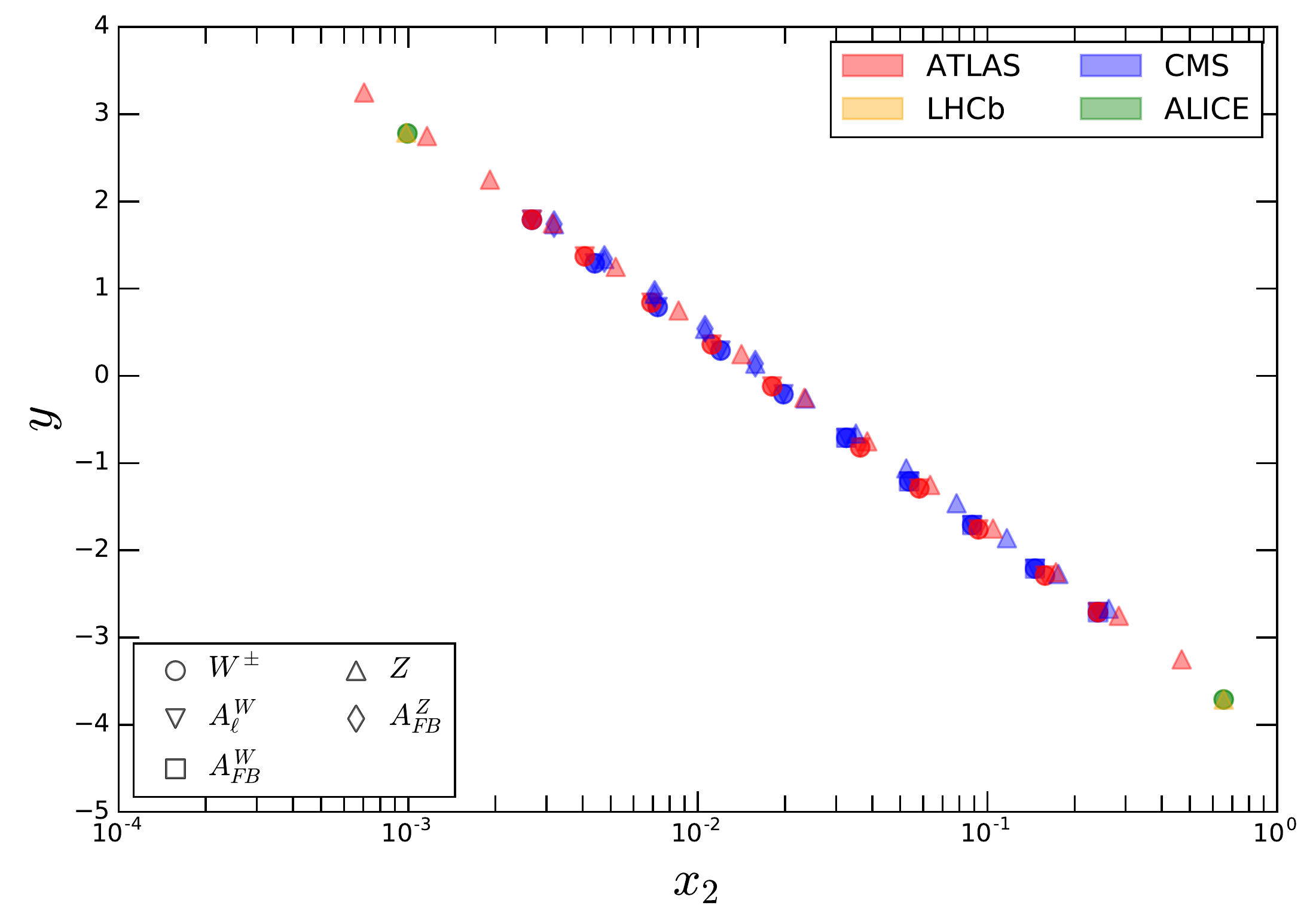}
\caption{Kinematic $x$--rapidity plane of lead covered by currently available LHC pPb $W/Z$ production data.}
\label{fig:y-x2}
\end{figure}

In both $W^{+}$ and $W^{-}$ cases we see a common pattern. The low rapidity
($y_{l^{\pm}}<0$) data are well described by the nPDFs, whereas when we go
towards larger rapidities ($y_{l^{\pm}}>0$) the deviations between data and
nPDF predictions grow.
It can be understood in the following way.
If we map the rapidity values to the $x$ of lead nucleus that
is probed%
    \footnote{This strictly holds only at leading order.}
we find that the negative rapidities correspond to moderate $x$ values
($\sim0.1$) and positive rapidities to the low $x$ values ($\sim3\times10^{-3}$),
see Fig.~\ref{fig:y-x2}. At the same time we know that the low-$x$ range of
nPDFs is unconstrained by the data currently used in the nPDF fits, so these
results come from an extrapolation of the larger $x$ region.

It is interesting to observe that a delayed shadowing (which shifts the
shadowing down to smaller $x$ values) would improve the comparison of the
data with the theory in the larger $y_{l^\pm}$ region; this type of behavior
was observed in the nuclear corrections extracted from the neutrino-DIS
charged current data~\cite{Kovarik:2010uv,Schienbein:2009kk,Nakamura:2016cnn,Kalantarians:2016jbj}.
Taking into account the errors from both the experimental data  and the
theoretical predictions, no definitive conclusions can be drawn at the present time.
Nonetheless, this data has the potential to strongly influence the nPDF fits,
especially in the small $x$ region. This will be even more pronounced with the new
data collected at the end of 2016, where nearly 10 times more statistics were recorded.

\section{Strange contribution}
\label{sec:strange}
In order to analyze our results more quantitatively, it is very useful
to look at PDF correlations. In particular, we are interested in assessing
the importance of the strange quark in our results. We will focus here on
the correlations between $W^+$ and $W^-$ cross sections, a more comprehensive
discussion including $Z$ cross section is presented in~\cite{Kusina:2016fxy}.
The correlations will be quantified by means of correlation cosine defined
in~\cite{Kusina:2016fxy,Nadolsky:2008zw}. In our figures they are plotted
as ellipses around central predictions for different PDFs.

\begin{figure}
\centering{}
\includegraphics[width=0.5\textwidth]{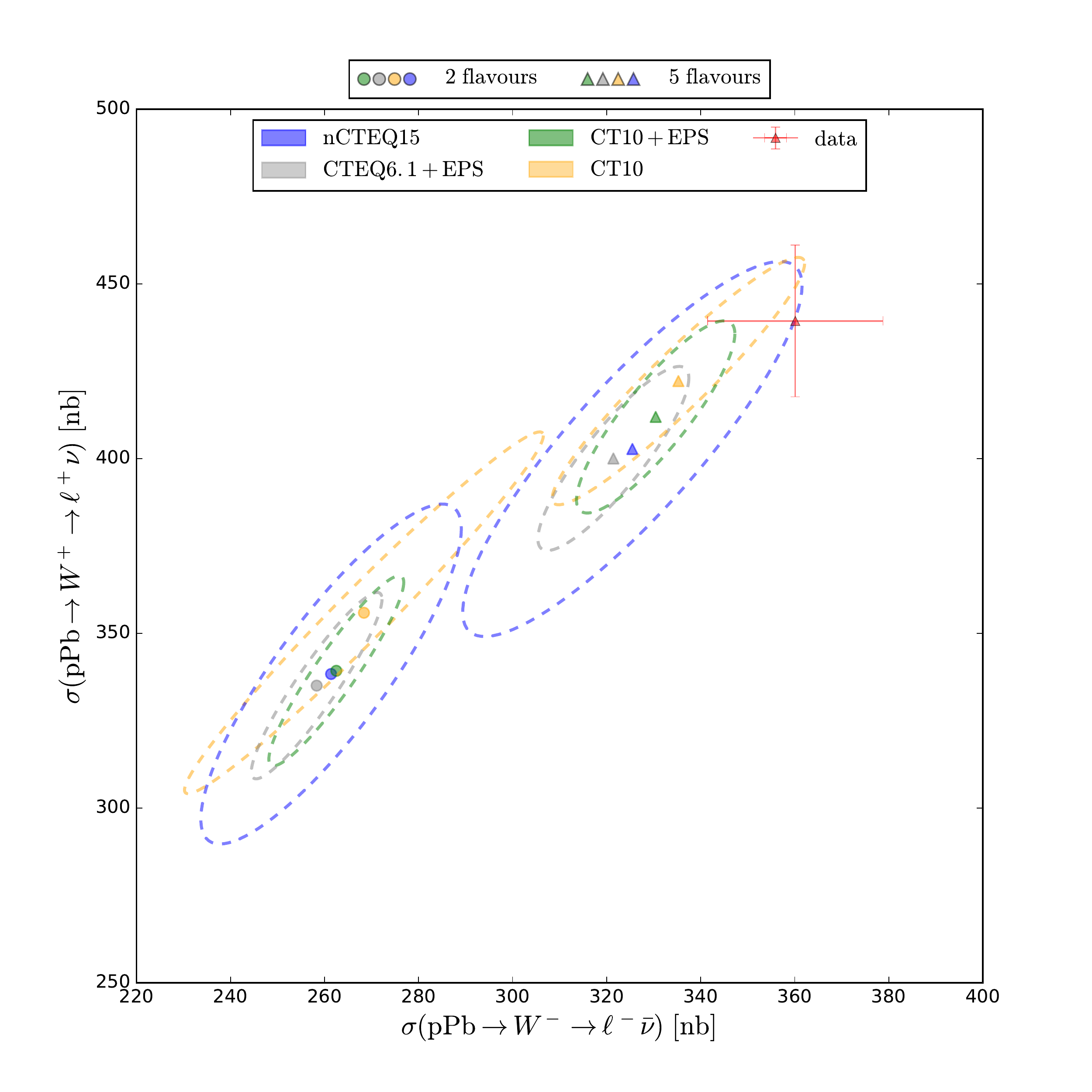}
\caption{Comparison of correlations between $W^+$ and $W^-$ cross sections for the case
when only one family of quarks $\{u,d\}$ is included and when all families are accounted for.}
\label{fig:correl_wp_wm_tot}
\end{figure}

\begin{figure}
\centering{}
\includegraphics[width=1.0\textwidth]{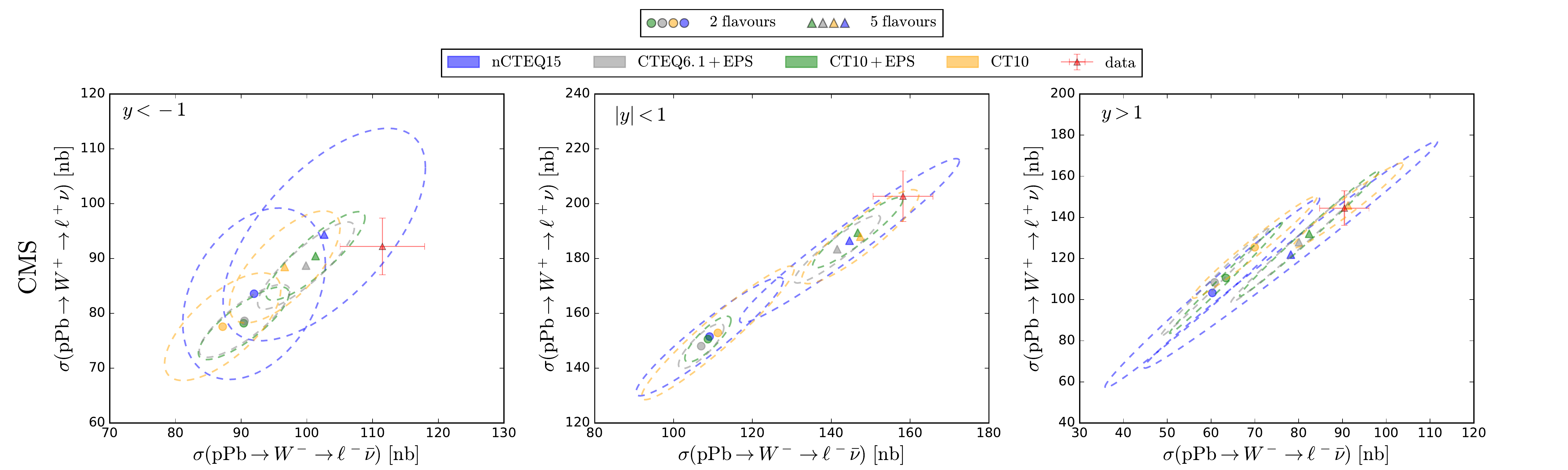}
\caption{Same as Fig.~\ref{fig:correl_wp_wm_tot} but divided into rapidity bins.}
\label{fig:correl_wp_wm_y}
\end{figure}

Fig.~\ref{fig:correl_wp_wm_tot} shows the correlations of the predicted $W^{+}$
and $W^{-}$ production cross sections for pPb collisions at the LHC in comparison
with the CMS measurements. The same result is displayed in
Fig.~\ref{fig:correl_wp_wm_y} but split into three different rapidity regions,
$y<-1,\ |y|<1,\ y>1$. For the proton side we always use the CT10 PDFs and for
the lead side we examine four cases:
i) nCTEQ15, 
ii) CT10,
iii) CT10 PDFs supplemented by the nuclear corrections from EPS09 (CT10+EPS09),
and iv) CTEQ6.1 proton PDFs supplemented by EPS09 nuclear corrections (CTEQ6.1+EPS09).
Additionally, to quantify the contribution of the strange quark we present also
results calculated using only 2 quark flavours (one family) $\{u,d\}$. In this way
we eliminate the contribution from the strange PDF (the c and b PDF contributions
are small).

The shift of the 2 flavor results compared to the 5 flavor results can be as
large as 30\% and reflects the large size of the strange contributions. 
The strange contributions to $W/Z$ boson production at the LHC are
substantial~\cite{Kusina:2012vh} and  are primarily responsible for the
observed differences among the nuclear results (nCTEQ15, EPS09+CT10, EPS09+CTEQ6.1). 
On the other hand, the observed differences between the 2 flavor proton CT10 and
the nuclear (nCTEQ15, EPS09) results accurately represent the nuclear corrections
associated with these quantities. 
Indeed, the nCTEQ15 and EPS09+CTEQ6.1 results are generally very close due to the
fact that the CTEQ6.1 and nCTEQ15 baseline PDFs are very similar.

As we review these correlation plots there are a number of general features
which we can identify.
As we move from negative $y$ to positive $y$ we move from high $x$ where the
nPDFs are well constrained to small $x$ where the nPDFs have large uncertainties
(still underestimated).
Thus, it is encouraging that at $y<-1$ we uniformly find the nuclear predictions
yield larger cross sections than the proton results (without nuclear corrections)
and thus lie closer to the LHC data.
Conversely, for   $y > 1$ we find the  {\mbox nuclear} predictions
yield smaller cross sections than the  proton results.
This situation suggests a number of possibilities. 

First, the large nPDF uncertainties in the small $x$ region could be improved
using the LHC data.

Second, the lower nPDF cross sections are partly due to the nuclear shadowing
in the small $x$ region; if, for example, this shadowing region were shifted to even
lower $x$ values, this would increase the nuclear results. 
Such a shift was  observed in Refs.~\cite{Kovarik:2010uv,Schienbein:2009kk,Nakamura:2016cnn}
using charged current neutrino-DIS data, and this would move the nuclear predictions
at $y>1$ toward the LHC data. 

Finally, we note that  measurements of the strange quark asymmetry~\cite{Mason:2007zz}
indicate that $s(x)\not=\bar{s}(x)$ which is unlike what is used in the current nPDFs;
this would influence the $W^\pm$ cross sections separately
as (at leading-order)
$W^+\sim \bar{s}c$
and $W^-\sim s\bar{c}$.
%
As the strange PDF has a large impact on the  $W^\pm/Z$  measurements,
this observation could provide incisive information on the
individual $s$ and $\bar{s}$ distributions.
%

\section{Impact of the data on nPDFs}
\label{sec:reweighting}
Ultimately, to see the impact of the LHC vector boson data on the nCTEQ15 PDFs
we will perform a new global analysis including these data. This work is ongoing
but in the meantime we try to estimate these effects by employing the reweighting
method~\cite{Giele:1998gw,Ball:2010gb,Paukkunen:2014zia,Sato:2013ika}.

In this exercise we use the Giele-Keller (GK) weight supplemented by
the tolerance criterion $T$ used in the nCTEQ15 fit
\begin{equation}
w_{k}=\frac{e^{-\frac{1}{2}\chi_{k}^{2}/T}}{\frac{1}{N_{\text{rep}}}
\sum_{i}^{N_{\text{rep}}}e^{-\frac{1}{2}\chi_{k}^{2}/T}},
\qquad
\chi_{k}^{2}=\sum_{j}^{N_{\text{data}}}\frac{(D_{j}-T_{j}^{k})^{2}}{\sigma_{j}^{2}},
\end{equation}
where $\chi_{k}^{2}$ represents the $\chi^{2}$ of the data sets considered
in the reweighting procedure for a given replica $k$.
This definition of the weight has been shown to reproduce the full Hessian
fit~\cite{Paukkunen:2014zia,Sato:2013ika}; as such it is an appropriate choice
for PDFs produced using the Hessian framework.
More details of the reweighting procedure can be found in our detailed
study~\cite{Kusina:2016fxy}.
Here we only note that after the reweighting, the PDF-dependent observables
and their errors can be computed as weighted sums
\begin{equation}
\begin{split}
\left<\ord\right>_{\text{new}} & =\frac{1}{N_{\text{rep}}}\sum_{k=1}^{N_{\text{rep}}}w_{k}\ord(f_{k}),
\\
\delta\left<\ord\right>_{\text{new}} & =\sqrt{\frac{1}{N_{\text{rep}}}
\sum_{k=1}^{N_{\text{rep}}}w_{k}\left(\ord(f_{k})-\left<\ord\right>\right)^{2}}.
\end{split}
\end{equation}

As an example, we consider the reweighting using the CMS $W^{\pm}$
production data from \ppb collisions~\cite{Khachatryan:2015hha}
(these data have the smallest uncertainties among the currently
available vector boson pPb data).
In this example we use rapidity distributions of charged leptons originating
from the decay of both $W$ bosons and we employ $N_{\text{r}ep}=10^{4}$
replicas.

\begin{figure}
\centering{}
\subfloat[$W^{+}$\label{fig:cms_wpm_pPb_comp_wp}]{
\includegraphics[width=0.48\textwidth]{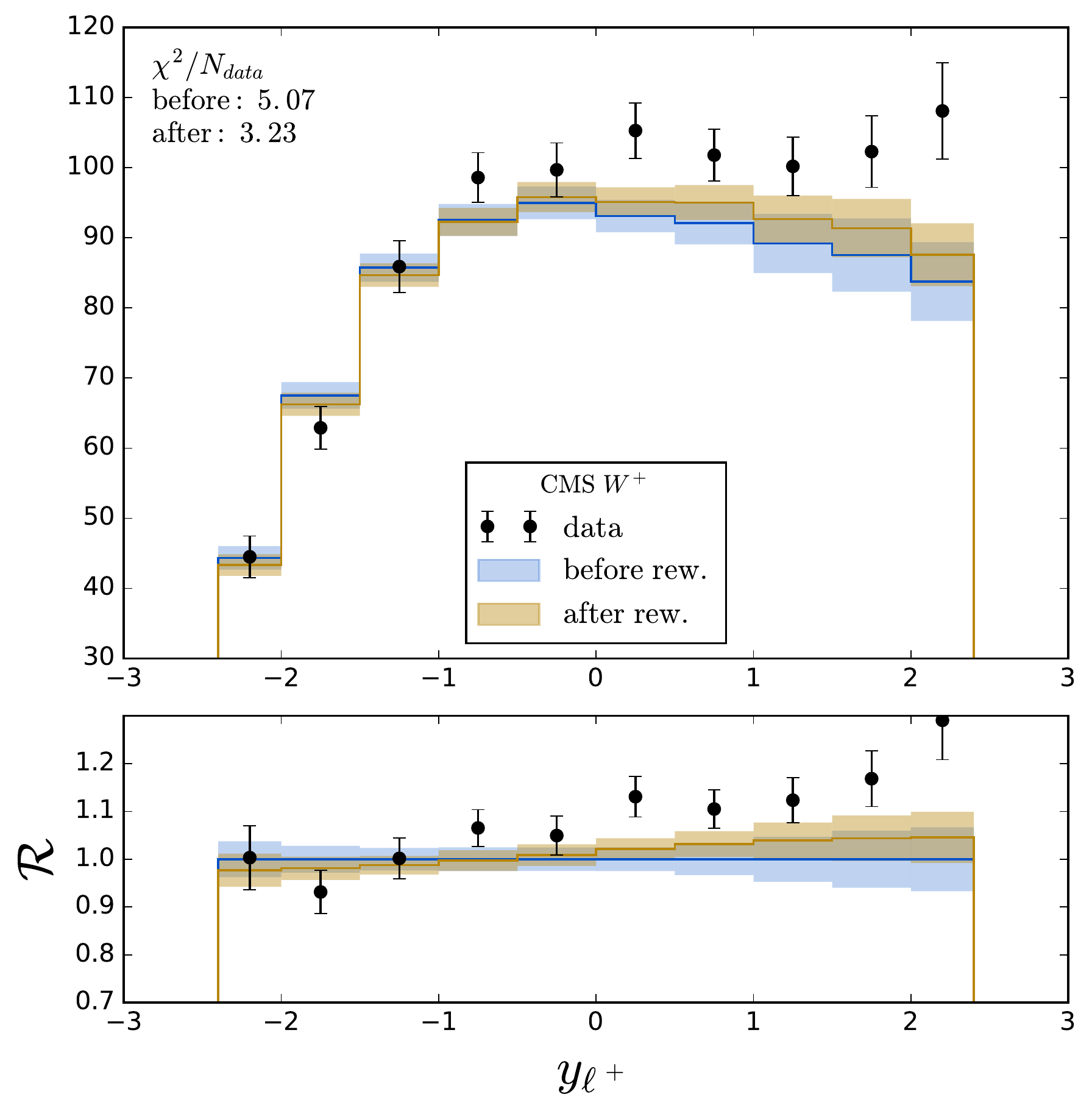}}
\hfil
\subfloat[$W^{+}$\label{fig:cms_wpm_pPb_comp_wp}]{
\includegraphics[width=0.48\textwidth]{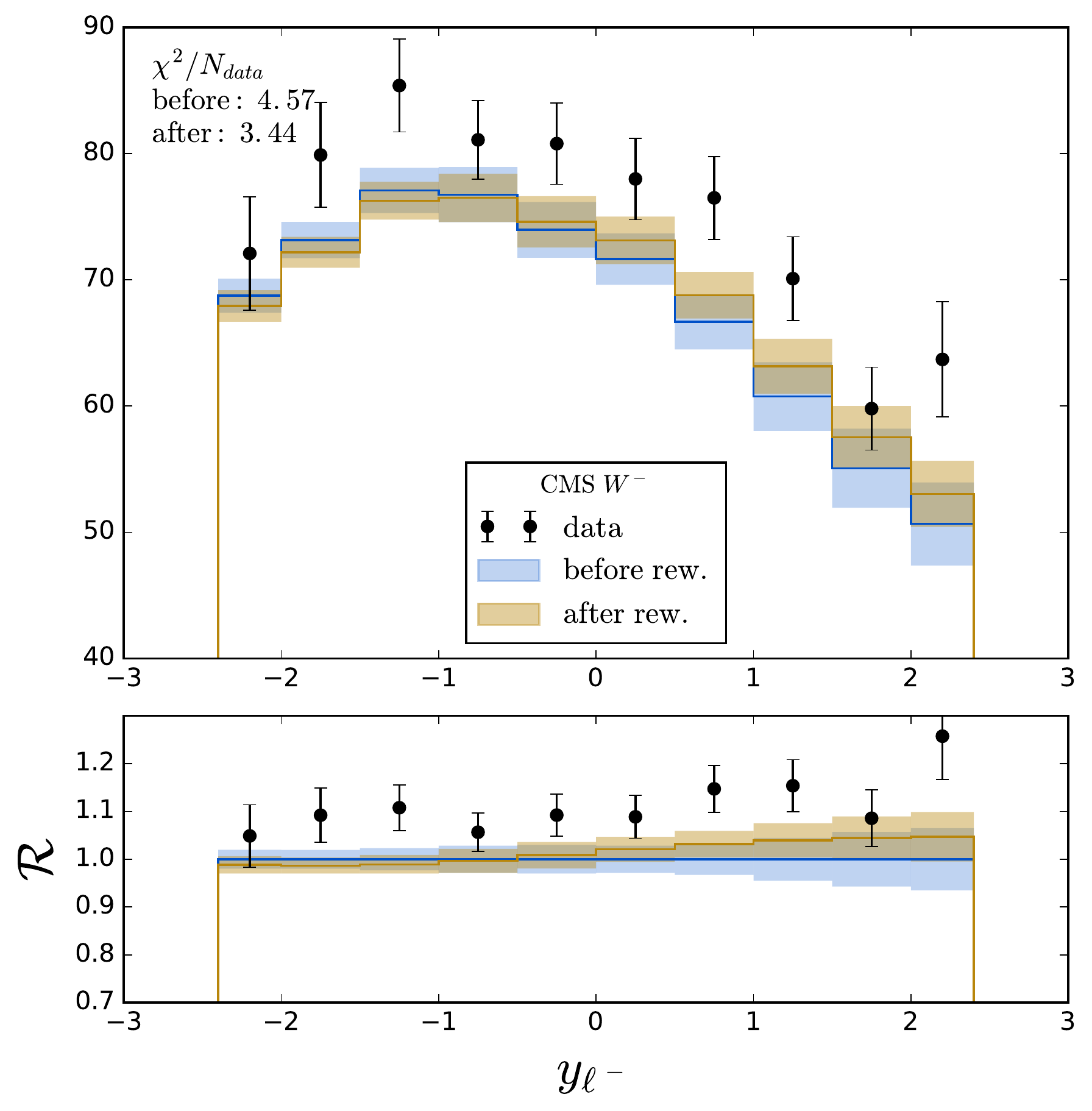}}
\caption{Comparison of data and theory before and after the reweighting using
CMS $W^{\pm}$ data for the nCTEQ15 PDFs.}
\label{fig:theo_after_before_cmsW}
\end{figure}

In Fig.~\ref{fig:theo_after_before_cmsW} we show the comparison of the data
to theory before and after the reweighting procedure.%
    \footnote{We note here the difference of PDF uncertainties compared
    to the plots presented in the previous sections. This is caused by the
    use of the 68\% c.l. errors compared to the standard nCTEQ15 90\% c.l.
    errors which were used earlier. This holds for all reweighting plots.}
As expected, we see that after the reweighting procedure
the description of the data is improved.
This is true for both the $W^{+}$ (left panel) and $W^{-}$ (right panel) cases.
We can quantify  the improvement of the fit by examining the
$\chi^{2}/N_{\text{data}}$ for the individual distributions.
For the $W^{+}$ case, the $\chi^{2}/N_{\text{data}}$ is improved from $5.07$ 
before  reweighting to $3.23$ after reweighting. Similarly, for
$W^{-}$ the $\chi^{2}/N_{\text{data}}$ is improved from $4.57$ to $3.44$. 
The amount of change due to the reweighting procedure should be proportional to the
experimental uncertainties of the incorporated data.
For $W^{\pm}$ production investigated here, the uncertainties are quite substantial, 
and the effects are compounded by the lack of correlated errors.

Still, even with the current data uncertainties we can see that the improvement
of the data description after the reweighting procedure is limited and
the resulting $\chi^{2}/N_{\text{data}}$ values are not satisfactory.
This is caused by considerably underestimated nPDF error bands, especially
in the positive rapidity region. As mentioned before, the low $x$ (large rapidity)
region of the current nPDFs is extrapolated as there are no constrains form data prior
to the LHC measurements. The underestimation of the errors is a result of too restrictive
parametrization form used in the nPDF analyses which, however, is necessary to obtain stable
fits.

This exercise shows that it is mandatory to release some of the assumptions and use more
flexible parameterizations to fully accommodate the vector boson LHC data in nPDF global fits.
This is also confirmed by the new EPPS16 analysis~\cite{Eskola:2016oht}.

\subsection{Strange contribution}
\label{sec:reweighting_str}
\begin{figure}
\centering{}
\subfloat[$W^{+}$\label{fig:cms_wpm_pPb_comp_wp}]{
\includegraphics[width=0.48\textwidth]{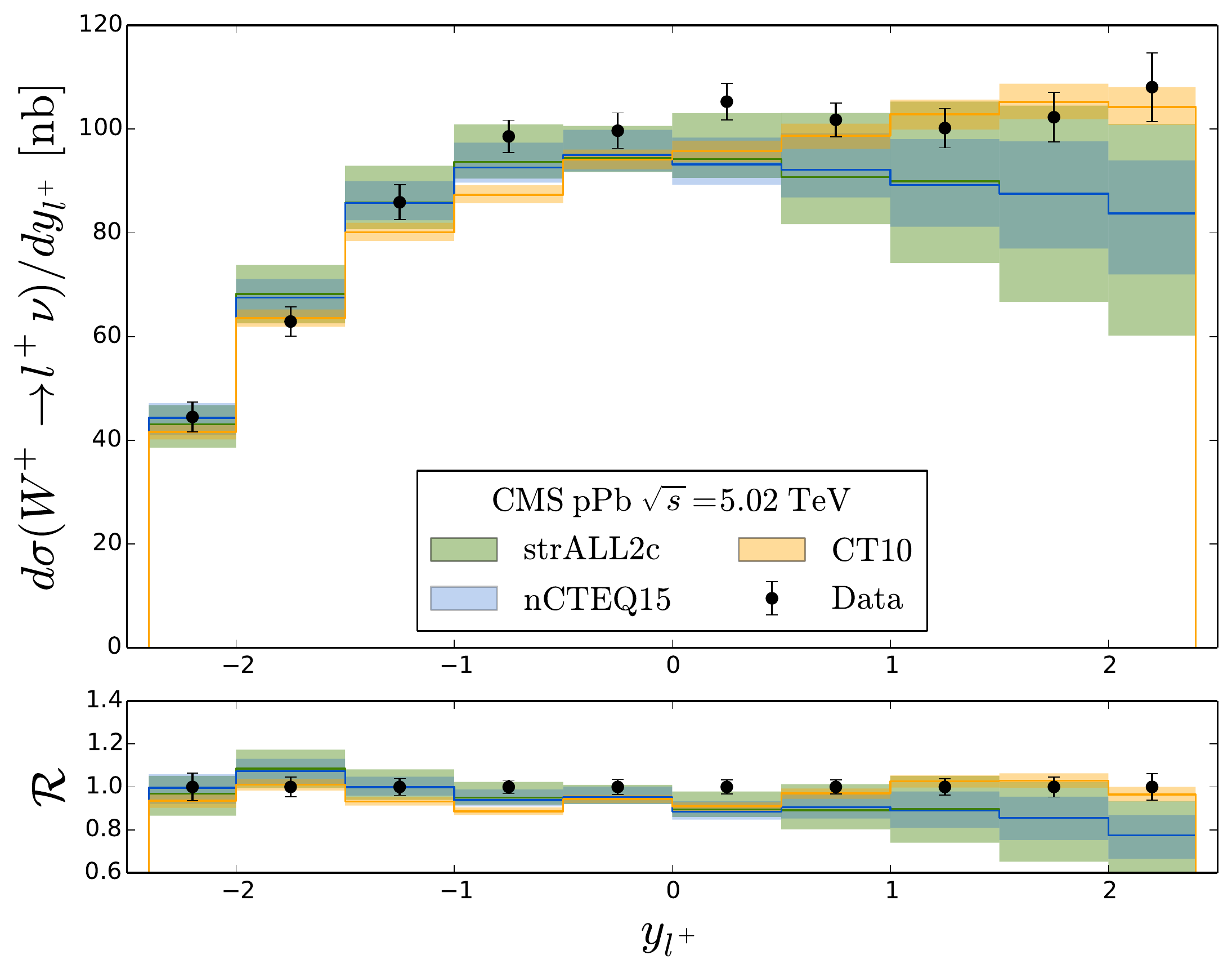}}
\hfil
\subfloat[$W^{+}$\label{fig:cms_wpm_pPb_comp_wp}]{
\includegraphics[width=0.48\textwidth]{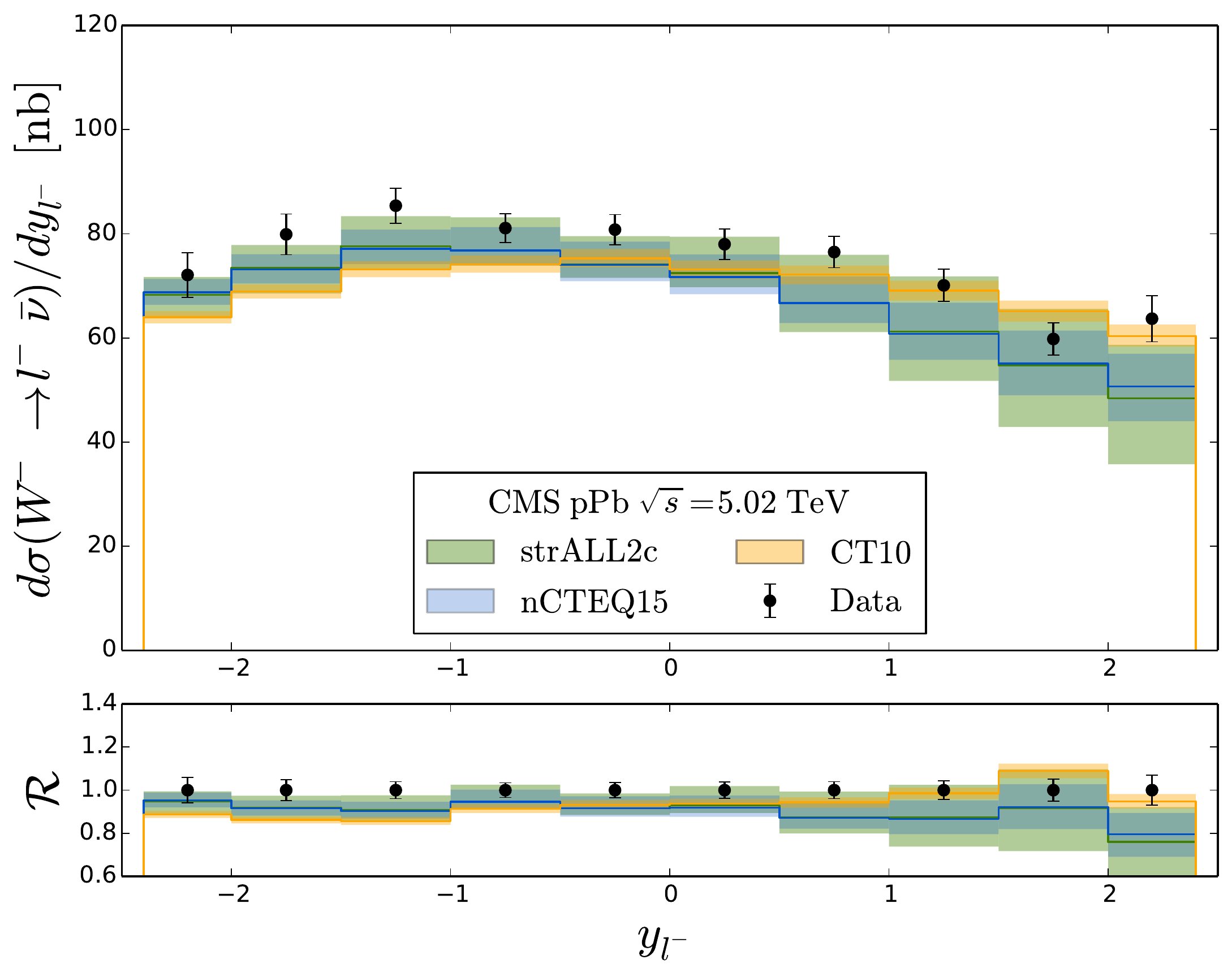}}
\caption{$W^{\pm}$ production in pPb collisions at the LHC from
CMS compared with predictions from nCTEQ15, CT10, and strALL2c
(nCTEQ15 with additional strange flexibility) PDFs.}
\label{fig:cmsWpm_str}
\end{figure}

We have shown that strange distribution is important for the $W/Z$ production at the LHC.
This clearly suggests that one of the reasons we have difficulties accommodating $W^{\pm}$
data in the current reweighting exercise
is the lack of proper estimates of strange distribution errors. Due to the lack of data the strange
PDF is not fitted in the nCTEQ15 and other nuclear analyses, but it is fixed to be proportional
to the light sea distribution $\bar{u}+\bar{d}$. We try to address this problem doing
a dedicated fit (referred to as strALL2c) where nCTEQ15 analysis is extended by including neutrino di-muon
data~\cite{Goncharov:2001qe}.%
    \footnote{The neutrino di-muon data are often used in proton PDF analyses to constrain
    strange distribution. They are however, rarely used in the nuclear PDF fits because
    of the unanswered question about the compatibility of the charge lepton and neutrino
    nuclear corrections.}
These data can put limited constraints on the strange allowing us to free some of the corresponding
fit parameters, and consequently provide more realistic error bars. This can be seen in
Fig.~\ref{fig:cmsWpm_str} where predictions for the CMS $W^{\pm}$ data for this new
fit with extra strange flexibility is compared to the original prediction for the nCTEQ15 PDFs.
We can see a substantial increase of the error bars in the positive rapidity (low $x$) region,
this is the extrapolation region, where nuclear PDF errors are underestimated to a large extent.

\begin{figure}
\centering{}
\subfloat[$W^{+}$\label{fig:cms_wpm_pPb_comp_wp}]{
\includegraphics[width=0.48\textwidth]{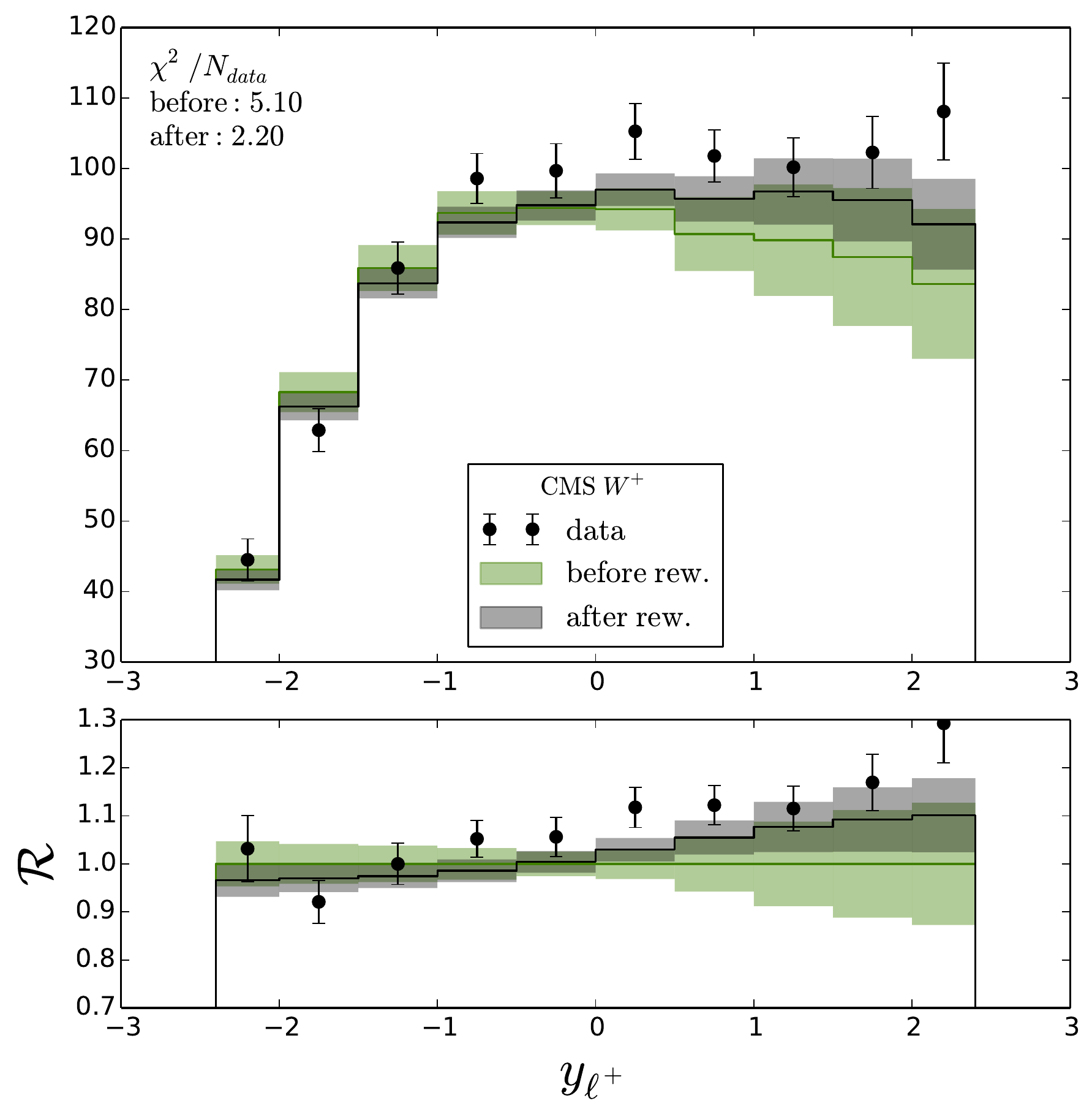}}
\hfil
\subfloat[$W^{+}$\label{fig:cms_wpm_pPb_comp_wp}]{
\includegraphics[width=0.48\textwidth]{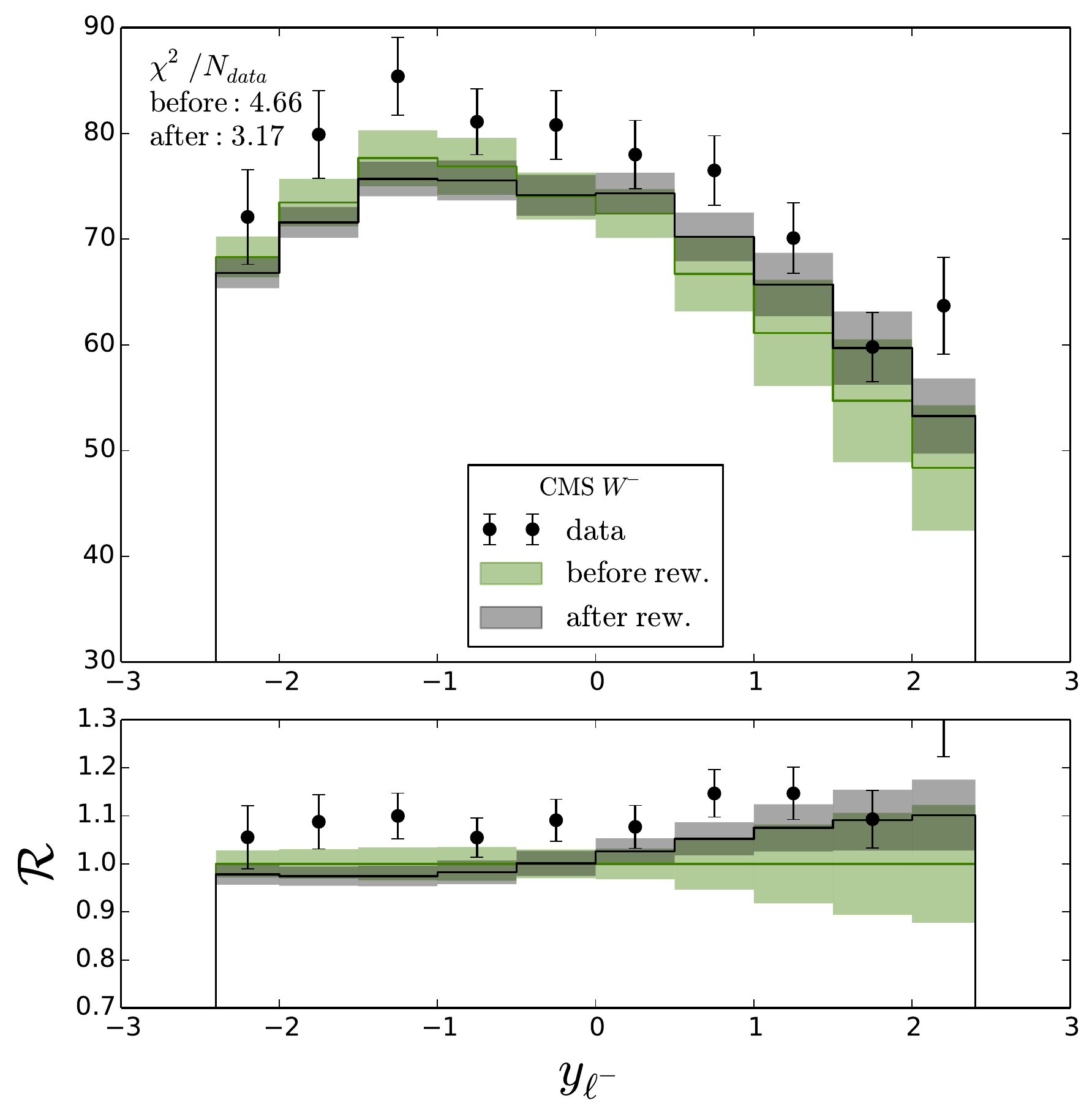}}
\caption{Comparison of data and theory before and after the reweighting using
CMS $W^{\pm}$ data for the strALL2c PDFs (nCTEQ15 with additional strange flexibility).}
\label{fig:theo_after_before_cmsW_str}
\end{figure}

To finish this exercise we perform a reweighting on the new strALL2c PDFs
to see if the additional flexibility allows to obtain more reliable results.
In Fig.~\ref{fig:theo_after_before_cmsW_str} we present the comparison of the data
to theory before and after the reweighting procedure using the strALL2c fit.
We can see that, indeed, the extra flexibility in strange distribution
(higher errors) allowed for more effective reweighting. The $\chi^{2}/N_{\text{data}}$
for $W^{+}$ case is now $2.20$ and for $W^{-}$ it is $3.17$. Especially in case of
the $W^{+}$ boson the improvement compared to the nCTEQ15 reweighting is substantial
(over 1 unit per data point).

This result shows that we are going in the right direction, however, the obtained 
$\chi^{2}/N_{\text{data}}$ are still relatively large and new fit with even more flexibility
is needed to properly incorporate the LHC vector boson data.

\section{Conclusions}
\label{sec:conclusions}

We have presented a study of vector boson production in lead collisions
at the LHC. These data are of particular interest for nPDF determinations.
A comparison with the LHC proton data provides a direct probe of nuclear
corrections for large $A$ values in a kinematic $\{x,Q^2\}$ range very different
from the nuclear corrections extracted from fixed-target measurements.

Our study has demonstrated the importance of the strange distribution for
the vector boson production at the LHC, possibly even pointing to a
nuclear strangeness asymmetry ($s(x)>\bar{s}(x)$).
More importantly, it showed that the currently used nuclear strange distributions
are not adequate and the characteristic underestimation of errors can cause problems
with the description of the LHC data.

This sensitivity to the strange distribution and heavier flavours can provide
important information on the nuclear flavor decomposition, which is invaluable
for a precise nPDF determination.

Intriguingly, the large rapidity $W/Z$ data seem to prefer nuclear PDFs
with no shadowing or delayed shadowing at small $x$, similar to what has been
observed in neutrino DIS.

\bibliographystyle{utphys}
\bibliography{refs}

\end{document}